%% file: Master_preprint.tex
\documentclass[aps,prd,preprint,showpacs]{revtex4-1}
\usepackage{def_symbol}
\usepackage[dvipdfmx]{color}
\usepackage{amsmath,amssymb}
\usepackage[dvipdfmx]{graphicx}

\begin{document}

\preprint{OU-HET-877}

\title{The Yang-Mills gradient flow and lattice effective action}


\author{Ryo Yamamura}
\email[]{ryamamura@het.phys.sci.osaka-u.ac.jp}
\affiliation{Department of Physics, Osaka University, \\
Toyonaka, Osaka 560-0043 Japan}



\begin{abstract}
Recently, the Yang-Mills gradient flow is found to be a useful concept not only in  
lattice simulations but also in continuous field theories.
Since its smearing property is similar to the Wilsoninan ``block spin transformation'',
there might be deeper connection between them.
In this work, we define the ``effective action'' which generates configurations at a finite flow time and
derive the exact differential equation to investigate the flow time dependence of the action.
Then Yang-Mills gradient flow can be regarded as the flow of the effective action.
We also propose the flow time dependent gradient,
where the differential equation becomes similar to the renormalization group equation.
We discuss a possibility to regard the time evolution of the effective action 
as the Wilsonian renormalization group flow.
\end{abstract}


\maketitle

\section{Introduction} 
\label{sec:Intro}
\input{Intro.tex}

\section{Exact differential equation for the effective action} 
\label{sec:WF}
\input{WF.tex}

\section{Extension to flow time dependent gradient flow} 
\label{sec:GF}
\input{GF.tex}

\section{Summary} 
\label{sec:Summary}
\input{Summary.tex}

\begin{acknowledgments}
I would like to express my deepest gratitude to Kengo Kikuchi for giving me a chance to consider this work.
I also thank Hidenori Fukaya, Tetsuya Onogi for fruitful discussions 
and careful reading of this manuscript,
and Akio Tomiya for useful comments.
This work is supported in part by the Grand-in-Aid of the Japanese Ministry of Education No. 15J01081.
\end{acknowledgments}

\bibliographystyle{apsrev4-1}
\bibliography{bib}

\appendix
\section{Calculation details for the derivation of Eqs.~\eqref{eq:DiffbetaGF0}-\eqref{eq:DiffbetaGF7}} 
\label{app:Relation}
\input{Appendix.tex}

\end{document}

%% file: Intro.tex
Recently, the Yang-Mills gradient flow (or the Wilson flow) 
\cite{Narayanan:2006rf, Luscher:2009eq, Luscher:2010iy, Luscher:2013vga} 
is found to be a useful concept in lattice simulations and widely applied to various issues.
Since it is regarded as a continuous stout smearing of the link variables,
effects coming from the lattice cut-off are reduced while the IR (or long-range) physics is kept unchanged.
Moreover, the fact that 
the smeared link variables look closer to the smooth renormalized fields
enables us to measure the topological charge and energy momentum tensor even on the lattice.

Not only in lattice simulations but also in continuous field theories,
the gradient flow shows an attractive property.
In the 4-dimensional pure Yang-Mills theory,
any correlation function in terms of the gauge fields at a finite flow time ($=t$)
are finite without additional renormalization, 
once the gauge coupling is renormalized suitably~\cite{Luscher:2011bx}.
The 2-dimensional $O(N)$ non-linear sigma model 
has also been studied and shown its renormalizability~\cite{Makino:2014sta,Aoki:2014dxa}.

Since the smearing property of the gradient flow is 
similar to the Wilsoninan ``block spin transformation'',
we want to reveal deeper relation between them,
or eventually construct the Wilsonian renormalization group (RG) using
good properties of the gradient flow.
One of the advantages of this approach is that 
the gradient flow is compatible with the gauge symmetry.
Thus there is a possibility to carry out the Wilsonian RG transformation 
in a gauge symmetric way, which is usually difficult in functional RG approaches with a momentum cut-off.

As a first step to achieve the above final goal,
it could be worth studying the ``effective action''  of the 4-dimensional lattice $SU(3)$ pure Yang-Mills theory,
which generates the configurations smeared by the Wilson flow.
This effective action is defined by changing the integration variables, 
\begin{align}
\label{eq:Intro}
e^{-S_t[V]}\equiv \int DU~\delta(V-\bar{V}_t[U])~e^{-\SW[U]},
\end{align}
where $\SW$ is the Wilson plaquette action and $\bar{V}_t[U]$ is the solution of the Wilson flow equation
with the initial condition $\bar{V}_0[U]=U$.

In our previous work~\cite{Kagimura:2015via},
the $t$-dependence of $S_t$ has been investigated by the lattice simulation.
$S_t$ is truncated so that it contains the Wilson plaquettes and the rectangular loops only and 
their corresponding couplings are determined by the demon method~\cite{Creutz:1983ra, Hasenbusch:1994ne}.
The result shows that the coupling of the plaquette grows 
while that of the rectangular tends to be negative with the flow time as the known improved actions~
\cite{Weisz:1982zw, Weisz:1983bn, Luscher:1984xn, Iwasaki:2011np, Iwasaki:1985we, deForcrand:1999bi}.
However the obtained trajectory in 2-coupling theory space 
travels in the opposite direction to the renormalization trajectory 
investigated by QCD-TARO collaboration~\cite{deForcrand:1999bi}.

In this work, we propose a different method to study the effective action defined by Eq.~\eqref{eq:Intro}.
We derive the exact differential equation for $S_t$ starting from the Wilson flow equation for the link variables.
The solution of this proposed equation tells us the $t$-dependence of couplings.
Unlike the previous work, we do not require implementing lattice simulations in our method.
Moreover, the differential equation is just a linear inhomogeneous first-order differential equation
and its solution can be obtained analytically.
For an application, we give a solution in a truncated 8-coupling theory space.
Note that the truncation of the action is just for a practical reason and 
our differential equation itself is exact without any truncation.
The result agrees with that of Ref.~\cite{Kagimura:2015via}
on the negativeness of the coupling of the rectangular.
On the other hand, the coupling of the plaquette shows the different behavior.
In contrast to the monotonic increase observed in Ref.~\cite{Kagimura:2015via},
it increases at a small $t$ region and tends to decrease at around $\sqrt{8t}\simeq1$ in our result.
We find that this difference can be understood as the difference of the truncation.


We also propose an extension of the differential equation to more generic cases:
it is not limited to the Wilson flow but can take any $t$-dependent lattice action for the gradient flow equation. 
For example, the ``trivializing action''~\cite{Luscher:2009eq}
which gives the vanishing $S_t$ at a fixed $t=t_0$ could be a good candidate.
In this work, $S_t$ itself is taken as a concrete example of the $t$-dependent action,
where the extended differential equation is quadratic in $S_t$ and becomes similar to the RG equation.
We also give a solution in the same truncated theory space as the case of the Wilson flow.
In this case, 
the coupling of the plaquette shows the similar behavior as explained in the previous paragraph
while the other 7 couplings rapidly grow with increasing the flow time.
Then we discuss a possibility to regard this time evolution as the Wilsonian RG flow.



The rest of our paper is organized as follows.
First, we define the exact differential equation for the effective action
and show its solution in a truncated theory space in Sect.~\ref{sec:WF}.
Then we extend the equation to the gradient flow with arbitrary $t$-dependent action in Sect.~\ref{sec:GF}.
The case where $S_t$ itself is chosen as a concrete example of 
the $t$-dependent action is considered in Sect.~\ref{subsec:Example}.
We discuss similarities and differences of the time evolution compared to the Wilsonian RG flow 
in Sect.~\ref{subsec:RelationRG}.
A summary is given in Sect.~\ref{sec:Summary}.

%% file: WF.tex
In this section, we derive the exact differential equation for the action
starting from the Wilson flow equation for the link variables.
The basic idea for the derivation has been proposed 
in the context of the exact renormalization group equation~\cite{Wegner:1974, Latorre:2000qc}.
In this work, we concentrate on the 4-dimensional lattice $SU(3)$ pure Yang-Mills theory,
but it is expected to be applied to various field theories.
We also 
give a solution of the obtained equation in truncated 8-coupling theory space. 

\subsection{
The effective action and its differential equation}
\label{subsec:Derive}
Let us define the effective action $S_t$ by the change of variables:
\begin{align}
\label{eq:defEA}
e^{-S_t[V]}\equiv \int DU~\delta(V-\bar{V}_t[U])~e^{-\SW[U]},
\end{align}
where $\SW[U]$ is the Wilson plaquette action,
\begin{equation}
\begin{split}
\label{eq:WilsonAction}
&\SW[U]=-\frac{1}{6}\hspace{0.5mm}\beta\hspace{-0.5mm}
\sum_{x,\mu\neq\nu}\hspace{-0.5mm}{\rm Tr}\hspace{0.5mm}W_{\mu\nu}(x)+{\rm const.},~~\beta=6/g^2_0,\\  
&W_{\mu\nu}(x)=U_\mu(x)U_\nu(x+\mu)U^{\dag}_\mu(x+\nu)U^\dag_\nu(x), 
\end{split}
\end{equation}
and $\bar{V}_t[U]=\{\bar{V}_\mu(x;t)\}$ is the solution of the Wilson flow equation,
\begin{equation}
\begin{split}
\label{eq:WFeq}
&\frac{d\bar{V}_\mu(x;t)}{dt}=
-g_0^2\{\partial_{x,\mu}S_{\rm W}[U]\}|_{U=\bar{V}_t}\bar{V}_\mu(x;t), \\ 
&\bar{V}_\mu(x;t)|_{t=0}=U_\mu(x). 
\end{split}
\end{equation}
Here we take $t$ to be a dimensionless parameter.
The differentiation with respect to the link variables on a differentiable function $f[U]$ is defined by
\begin{equation}
\begin{split}
\label{eq:DiffLink}
&\pa^a_{x,\mu}f[U]=\frac{d}{ds}f[U^s]\bigg|_{s=0}, \\[0.2cm]
&U^s_\nu(y)=
\bigg\{
\begin{array}{ll}
e^{sT^a}U_\mu(x) &~{\rm if}~(y,\nu)=(x,\mu),\\
U_\nu(y) &~{\rm otherwise},
\end{array}
\end{split}
\end{equation}
and $\pa_{x,\mu}=T^a\pa^a_{x,\mu}$,
where $T^a$ are the anti-hermitian $SU(3)$ generators whose normalization is given by
\begin{align}
\label{eq:Trace}
\Tr(T^aT^b)=-\frac{1}{2}\delta^{ab}.
\end{align}
With Eq.~\eqref{eq:DiffLink} and Eq.~\eqref{eq:Trace},
an infinitesimal change of $f[U]$ is 
\begin{align}
\label{eq:InfChange}
d\hspace{0.1mm}f[U]=-2\sum_{x,\mu}\Tr[d\hspace{0.1mm}U_\mu(x)U^{-1}_\mu(x)T^a]~\pa^a_{x,\mu}f[U].
\end{align}

To derive the differential equation for $S_t$, 
we differentiate both hand sides of Eq.~\eqref{eq:defEA} with respect to $t$,
\begin{align}
\label{eq:DiffWitht}
-\frac{dS_t[V]}{dt}e^{-S_t[V]}
=&~\int \hspace{-1mm}DU\frac{d}{dt}[\delta(V-\bar{V}_t[U])]\hspace{1mm}e^{-\SW[U]}.
\end{align}
The differentiation of the delta function 
in Eq.~\eqref{eq:DiffWitht} is evaluated as
\begin{align}
\frac{d}{dt}\delta(V-\bar{V}_t[U])
\label{eq:Diffdelta}
=&~g^2_0\sum_{x,\mu}\bar{\pa}^a_{x,\mu}S_{\rm W}[\bar{V}_t]~\pa^a_{x,\mu}[\delta(V-\bar{V}_t[U])],
\end{align}
where $\pa^a_{x,\mu}$ and $\bar{\pa}^a_{x,\mu}$ denote the differentiation 
with respect to $V_\mu(x)$ and $\bar{V}_\mu(x;t)$ respectively.
Substituting Eq.~\eqref{eq:Diffdelta} for Eq.~\eqref{eq:DiffWitht}, we finally obtain
\begin{align}
-\frac{dS_t[V]}{dt}e^{-S_t[V]}
\label{eq:Mastermae}
=&~g^2_0\sum_{x,\mu}\pa^a_{x,\mu}\bigg[\pa^a_{x,\mu}S_{\rm W}[V]e^{-S_t[V]}\bigg],
\end{align}
or equivalently,
\begin{align}
\label{eq:MasterEqWF}
\frac{dS_t[V]}{dt}=
g^2_0\sum_{x,\mu}\{\pa^a_{x,\mu}S_t[V]\pa^a_{x,\mu}S_{\rm W}[V]-(\pa^a_{x,\mu})^2\SW[V]\}.
\end{align}
Note that this equation is independent of $g_0^2$ since $\SW\propto 1/g^2_0$ and 
the overall $g^2_0$ on the right hand side of Eq.~\eqref{eq:MasterEqWF} is canceled.

\subsection{Solution in truncated theory space}
\label{subsec:SolutionWF}
\begin{figure}[tbhp]
\centering
\includegraphics[width=10cm]{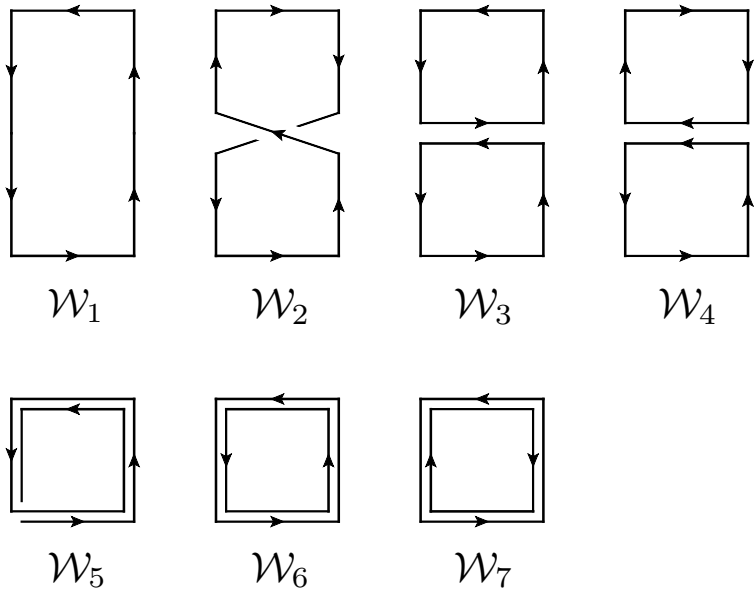}
\caption{Classes of Wilson loops and pairs of them in the truncated $S_t$.
The loops $5,6,7$ lie on a single plaquette of the lattice. 
The other loops occupy two plaquettes which can lie in a plane or be at right angles in three dimensions.}
\label{fig:7WL}
\end{figure}

We demonstrate computing $S_t$ by solving Eq.~\eqref{eq:MasterEqWF} in this subsection.
To this end, we truncate $S_t$, as is often the case in studies of a functional renormalization group.
Note that this truncation is just for a practical reason and 
Eq.~\eqref{eq:MasterEqWF} itself is exact without any truncation.
In the following, we take the truncation so that $S_t$ is in the 8-coupling theory space:
\begin{align}
\label{eq:TrunAction}
S_t[V]=-\frac{1}{6}\sum_{i=0}^7\beta_i(t)\mathcal{W}_i[V]+({\rm independent~of~}V),
\end{align}
where each $\mathcal{W}_i$ is defined by the sum of the trace of the associated Wilson loops over the whole lattice,
\begin{equation}
\begin{split}
&\mathcal{W}_i=\sum_{C\in\mathcal{C}_i}\Tr\{U(C)\}~~{\rm if}~i=0,1,2,5, \\
&\mathcal{W}_i=\hspace{-3mm}
\sum_{\{C,C'\}\in\mathcal{C}_i}\hspace{-3mm}\Tr\{U(C)\}\Tr\{U(C')\}~~{\rm if}~i=3,4,6,7.
\end{split}
\end{equation}
Here $U(C)$ denotes the the ordered product of the link variables along the loop $C$.
The Wilson plaquette is associated with $\mathcal{C}_0$. 
The other classes $\mathcal{C}_{i\geq 1}$ 
of Wilson loops and pairs of them 
are constructed by the contraction of two Wilson plaquettes with a common link variable (see Fig.~\ref{fig:7WL}).
Hereafter we omit the $V$-independent term in $S_t$.
We also take the initial condition as $S_0=\SW$, specifically,
\begin{equation}
\begin{split}
\label{eq:InitBeta}
&\beta_0(0)=\beta, \\
&\beta_i(0) = 0~~{\rm if}~i=1,\dots,7,
\end{split}
\end{equation}
for simplicity.

Note that the truncation is reasonable in a small $t$ regime,
since contributions from the loops which occupy larger area of the lattice is not negligible
in a large $t$ regime.
In fact, the solution under the above truncation scheme are same as the exact one up to $O(t^2)$.
For higher orders in $t$,
the number of the required Wilson loops grows factorially
and analytic calculation becomes harder.

Now let us rewrite Eq.~\eqref{eq:MasterEqWF} as the differential equations for $\beta_i(t)$.
The left hand side of Eq.~\eqref{eq:MasterEqWF} becomes
\begin{align}
\label{eq:LHS}
\frac{dS_t[V]}{dt}=-\frac{1}{6}\sum_{i=0}^{7}\frac{d\beta_i(t)}{dt}\mathcal{W}_i[V],
\end{align}
where the differentiation is performed at constant $V$.

For computing the right hand side of Eq.~\eqref{eq:MasterEqWF},
we make use of an identity for the $SU(3)$ generators,
\begin{align}
\label{eq:Identity}
(T^a)_{\alpha\beta}(T^a)_{\gamma\delta}
=-\frac{1}{2}\left(\delta_{\alpha\delta}\delta_{\beta\gamma}
-\frac{1}{3}\delta_{\alpha\beta}\delta_{\gamma\delta}\right).
\end{align}
Some algebraic manipulations yield
\begin{align}
&~\sum_{x,\mu}\pa^a_{x,\mu}\mathcal{W}_0\hspace{0.5mm}\pa^a_{x,\mu}\mathcal{W}_0
=\mathcal{W}_1-\mathcal{W}_2-\tfrac{1}{3}\mathcal{W}_3+\tfrac{1}{3}\mathcal{W}_4 \notag \\[-3mm]
\label{eq:W0}
&~~~~~~~~~~~~~~~~~~~~~~~~~~
-2\mathcal{W}_5+\tfrac{2}{3}\mathcal{W}_6-\tfrac{4}{3}\mathcal{W}_7, \\[1em]
\label{eq:W1}
&~\sum_{x,\mu}\pa^a_{x,\mu}\mathcal{W}_1\hspace{0.5mm}\pa^a_{x,\mu}\mathcal{W}_0
=30\mathcal{W}_0+\cdots, \\
\label{eq:W2}
&~\sum_{x,\mu}\pa^a_{x,\mu}\mathcal{W}_2\hspace{0.5mm}\pa^a_{x,\mu}\mathcal{W}_0
=50\mathcal{W}_0+\cdots, \\
\label{eq:W3}
&~\sum_{x,\mu}\pa^a_{x,\mu}\mathcal{W}_3\hspace{0.5mm}\pa^a_{x,\mu}\mathcal{W}_0
=120\mathcal{W}_0+\cdots, \\
\label{eq:W4}
&~\sum_{x,\mu}\pa^a_{x,\mu}\mathcal{W}_4\hspace{0.5mm}\pa^a_{x,\mu}\mathcal{W}_0
=120\mathcal{W}_0+\cdots, \\
\label{eq:W5}
&~\sum_{x,\mu}\pa^a_{x,\mu}\mathcal{W}_5\hspace{0.5mm}\pa^a_{x,\mu}\mathcal{W}_0
=4\mathcal{W}_0+\cdots, \\
\label{eq:W6}
&~\sum_{x,\mu}\pa^a_{x,\mu}\mathcal{W}_6\hspace{0.5mm}\pa^a_{x,\mu}\mathcal{W}_0
=12\mathcal{W}_0+\cdots, \\
\label{eq:W7}
&~\sum_{x,\mu}\pa^a_{x,\mu}\mathcal{W}_7\hspace{0.5mm}\pa^a_{x,\mu}\mathcal{W}_0
=6\mathcal{W}_0+\cdots, \\
\label{eq:W0^2}
&~\sum_{x,\mu}(\pa^a_{x,\mu})^2\mathcal{W}_0=-\tfrac{16}{3}\mathcal{W}_0,
\end{align}
up to an irrelevant constant in $V$.
Omitted part in Eqs.~\eqref{eq:W1}-\eqref{eq:W7} does not contain $\mathcal{W}_{1\leq i\leq7}$.

Comparing the coefficient of $\mathcal{W}_i$ on the both hand sides of Eq.~\eqref{eq:MasterEqWF}, 
we finally obtain 
\begin{align}
\label{eq:DiffbetaWF}
\frac{d\beta_i(t)}{dt}=\sum_{j=0}^7M_{ij}\beta_j(t)+32\delta_{i0},
\end{align}
where $M_{ij}$ is the $8\times 8$ real matrix given by 
\begin{align}
\label{eq:MatTru}
M_{ij}=
\left(
\begin{array}{cccccccc}
0 & -30 & -50 & -120 & -120 & -4 & -12 & -6 \\
-1 & 0 & 0 & 0 & 0 & 0 & 0 & 0 \\
1 & 0 & 0 & 0 & 0 & 0 & 0 & 0 \\
\frac{1}{3} & 0 & 0 & 0 & 0 & 0 & 0 & 0 \\
-\frac{1}{3} & 0 & 0 & 0 & 0 & 0 & 0 & 0 \\
2 & 0 & 0 & 0 & 0 & 0 & 0 & 0 \\
-\frac{2}{3} & 0 & 0 & 0 & 0 & 0 & 0 & 0 \\
\frac{4}{3} & 0 & 0 & 0 & 0 & 0 & 0 & 0 
\end{array}
\right),
\end{align}
within our truncation scheme.
Note that since the right hand side of Eq.~\eqref{eq:MasterEqWF} is linear in $S_t$,
the differential equation of $\beta_i(t)$ always forms as Eq.~\eqref{eq:DiffbetaWF} with any truncation.
Eq.~\eqref{eq:DiffbetaWF} is just a
linear inhomogeneous first-order differential equation and easily solved analytically.

\begin{figure}[tbhp]
\centering
\hspace{2cm}\includegraphics[width=12cm]{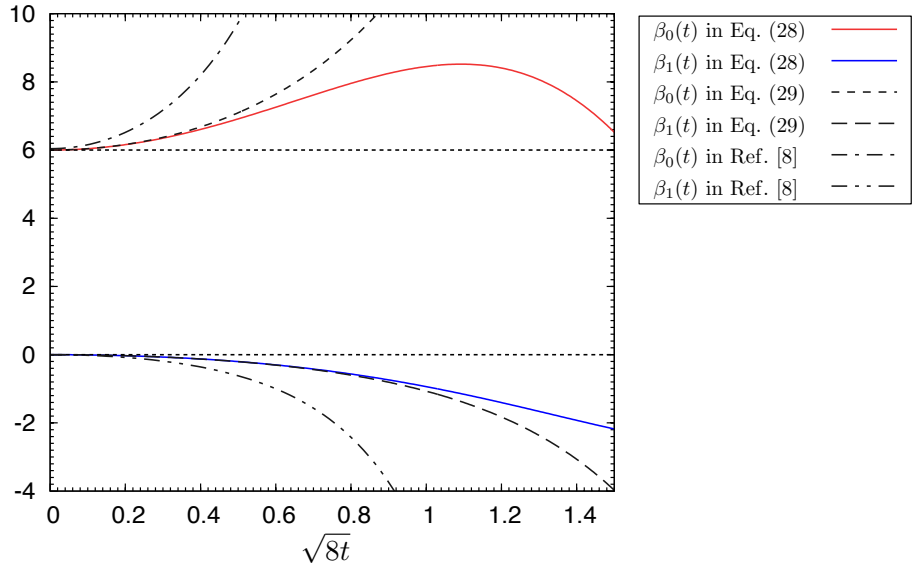}
\caption{
$\beta_0(t)$ and $\beta_1(t)$ as functions of $\sqrt{8t}$.
The value of $\beta$ which is the initial value of $\beta_0(t)$ 
is taken to be 6 as same as that in Ref.~\cite{Kagimura:2015via}.
The solid lines represent $\beta_0(t)$ and $\beta_1(t)=-\gamma(t)$ in Eq.~\eqref{eq:Sols},
the dashed lines represent those in Eq.~\eqref{eq:2WLTrun} and 
the dot-dashed lines represent those determined in Ref.~\cite{Kagimura:2015via} by the lattice simulation.
}
\label{fig:SolWF}
\end{figure}

The solutions of $\beta_i(t)$ with Eq.~\eqref{eq:InitBeta} and Eq.~\eqref{eq:MatTru} are given by
\begin{equation}
\begin{split}
\label{eq:Sols}
&\beta_0(t)=\beta\cos\left(2\sqrt{7}\hspace{0.3mm}t\right)
+\frac{16}{\sqrt{7}}\sin\left(2\sqrt{7}\hspace{0.3mm}t\right),  \\
&\beta_i(t)=M_{i0}\gamma(t),~~{\rm for~~}i=1,\dots, 7, \\
&\gamma(t)=\frac{8}{7}\left[1-\cos\left(2\sqrt{7}\hspace{0.3mm}t\right)\right]
+\frac{\beta}{2\sqrt{7}}\sin\left(2\sqrt{7}\hspace{0.3mm}t\right).
\end{split}
\end{equation}
We plot these solutions as the functions of $\sqrt{8t}$ in Fig.~\ref{fig:SolWF}.
$\beta=6$ is taken for the initial condition.
Note that $\beta_1(t)=M_{10}\gamma(t)=-\gamma(t)$ and 
$\beta_{i\geq2}(t)$ are proportional to $\gamma(t)$.
Two dot-dashed lines in Fig.~\ref{fig:SolWF} represent 
$\beta_0(t)$ and $\beta_1(t)$ obtained by the numerical fit in Ref.~\cite{Kagimura:2015via} respectively,
where $S_t$ is truncated so that it has the Wilson plaquettes and rectangular loops only.

From Fig.~\ref{fig:SolWF}, the two approaches agree on the negativeness of $\beta_1(t)$ at $t>0$,
which has been already reported by Ref.~\cite{Kagimura:2015via},
however the flow time dependence is clearly different.
This can be understood as the difference of the truncation of $S_t$, 
since we obtain 
\begin{equation}
\begin{split}
\label{eq:2WLTrun}
&\beta_0(t)=\beta\cosh\left(\sqrt{30}\hspace{0.3mm}t\right)
+\frac{16\sqrt{30}}{15}\sinh\left(\sqrt{30}\hspace{0.3mm}t\right), \\
&\beta_1(t)=-\frac{16}{15}\left[\cosh\left(\sqrt{30}\hspace{0.3mm}t\right)-1\right]
-\frac{\sqrt{30}}{30}\beta\sinh\left(\sqrt{30}\hspace{0.3mm}t\right),
\end{split}
\end{equation}
when we set $\beta_{i\geq2}(t)$
to be 0 in Eq.~\eqref{eq:DiffbetaWF}.
These solutions are also plotted as two dashed lines in Fig.~\ref{fig:SolWF}. 
The similar exponential dependence in $t$ has been observed in Ref.~\cite{Kagimura:2015via},
although the value of the exponent still does not coincide.

For our solution, the flow time dependence of $\beta_0(t)$ looks interesting.
Since the expectation value of the plaquette at a finite $t$ is considered to be 
larger than that of the plaquette at $t=0$, 
we naively expect the coupling of the plaquette to increase monotonically with $t$.
On the other hand, our result shows that 
it increases at a small $t$ region due to the second term 
on the right hand side of Eq.~\eqref{eq:DiffbetaWF},
then turns to decrease at around $\sqrt{8t}\simeq1$.
This apparently looks inconsistent with our naive guess.
However we may not discuss the value of the plaquette based only on $\beta_0(t)$
because the inverse of the squared ``effective gauge coupling'' 
is not given by $\beta_0(t)$ but given by a suitable linear combination of $\beta_i(t)$.
Algebraically, the second derivative of $\beta_0(t)$ with respect to $t$ satisfies
\begin{align}
\label{eq:SecondDiv}
\frac{d^2\hspace{-0.5mm}\beta_0(t)}{dt^2}=M^2_{00}\beta_{0},~~M^2_{00}=-28<0,
\end{align}
in this truncation, which causes such a behavior of $\beta_0(t)$.
Note that $d^2\hspace{-0.3mm}\beta_0(t)/dt^2<0$ is just a consequence of the algebraic manipulation and 
truncation effects on these solutions should be kept in mind.

%% file: GF.tex
In this section, 
we consider the extension of Eq.~\eqref{eq:MasterEqWF} to more generic cases.
Specifically, the flow equation is not limited to the Wilson flow equation (Eq.~\eqref{eq:WFeq})
but can be taken as a flow equation with an arbitrary chosen action.
For example, the ``trivializing action''~\cite{Luscher:2009eq} 
which gives the vanishing $S_t$ at a fixed $t=t_0$ could be a good candidate.
$S_t$ itself can also be chosen as the seeding action generating $S_{t+dt}$
as similar as a conventional RG transformation.

\subsection{Extension of Eq.~\eqref{eq:MasterEqWF}}
\label{subsec:Extension}
To extend Eq.~\eqref{eq:MasterEqWF} to more generic cases,
let us consider an infinitesimal change of the flow time at $t$ as follows,
\begin{align}
\label{eq:InfChange}
e^{-S_{t+dt}[V']}=\int DV~\delta(V'-\bar{V}_{dt}[V])~e^{-S_t[V]},
\end{align}
Suppose that $\bar{V}_{dt}[V]$ is given by 
\begin{align}
\label{eq:}
\bar{V}_\mu(x;dt)[V] = V_\mu(x) -\{\pa_{x,\mu}\mathcal{S}_t[V]\}V_\mu(x)dt + O((dt)^2),
\end{align}
where $\mathcal{S}_t[V]$ is an arbitrary $t$-dependent action
generating the flow of the link variables from $t$ to $t+dt$.
Then by comparing the $O(dt)$ terms on both hand sides of Eq.~\eqref{eq:InfChange},
we obtain the flow time dependent gradient flow equation for $S_t$:
\begin{align}
\label{eq:MasterEqGF}
\frac{dS_t[V]}{dt}=
\sum_{x,\mu}\{\pa^a_{x,\mu}S_t[V]\pa^a_{x,\mu}\mathcal{S}_t[V]
-(\pa^a_{x,\mu})^2\mathcal{S}_t[V]\},
\end{align}
where an initial action can be any lattice action which consists of Wilson loops and products of them.
When we choose $\mathcal{S}_t=g^2_0\SW$, 
Eq.~\eqref{eq:MasterEqGF} is equivalent to Eq.~\eqref{eq:MasterEqWF}
in which the flow of $S_t$ is generated by the Wilson flow. 

\subsection{A concrete example: $\mathcal{S}_t=g^2_0S_t$}
\label{subsec:Example}
In this subsection, we analyze Eq.~\eqref{eq:MasterEqGF} in the case of $\mathcal{S}_t=g^2_0S_t$, namely 
\begin{align}
\label{eq:MasterEqGFEx}
\frac{dS_t[V]}{dt}=
g^2_0\sum_{x,\mu}\{\pa^a_{x,\mu}S_t[V]\pa^a_{x,\mu}S_t[V]
-(\pa^a_{x,\mu})^2S_t[V]\},
\end{align}
with the same truncation (Eq.~\eqref{eq:TrunAction} and Fig.~\ref{fig:7WL}) and 
initial condition (Eq.~\eqref{eq:InitBeta}) of $S_t$ as in Sect.~\ref{subsec:SolutionWF}.

The computation is almost same as in the previous section.
For this case, we additionally need the relations in App.~\ref{app:Relation}
in order to derive the following differential equation for $\beta_i(t)$,
\begin{align}
&~\frac{d\beta_0(t)}{dt}=
-\frac{\beta_0(t)}{\beta}[60\beta_1(t)+100\beta_2(t)+240\beta_3(t)+240\beta_4(t) \notag \\
\label{eq:DiffbetaGF0}
&~~~~~~~~~~~~~~~~~~~~~~+8\beta_5(t)+24\beta_6(t)+12\beta_7(t)-32], \\
\label{eq:DiffbetaGF1}
&~\frac{d\beta_1(t)}{dt}=-\frac{1}{\beta}\left[\beta^2_0(t)+12\beta^2_1(t)+28\beta^2_2(t)+20\beta_2(t)\beta_5(t)
-48\beta_1(t)+6\beta_3(t)\right], \\
\label{eq:DiffbetaGF2}
&~\frac{d\beta_2(t)}{dt}=-\frac{1}{\beta}\left[-\beta^2_0(t)+24\beta_1(t)\beta_2(t)+12\beta_1(t)\beta_5(t)
-62\beta_2(t)-6\beta_4(t)\right], \\
&~\frac{d\beta_3(t)}{dt}=-\frac{1}{\beta}\left[-\tfrac{1}{3}\beta^2_0(t)+24\beta_1(t)\beta_3(t)
+6\beta_1(t)\beta_7(t)+40\beta_2(t)\beta_4(t)+20\beta_2(t)\beta_6(t)\right. \notag \\
&~~~~~~~~~~~~~~~~~~~+48\beta^2_3(t)+24\beta_3(t)\beta_7(t)+48\beta^2_4(t)+16\beta_4(t)\beta_5(t) 
+48\beta_4(t)\beta_6(t) \notag \\[1mm]
\label{eq:DiffbetaGF3}
&~~~~~~~~~~~~~~~~~~~\left.-~66\beta_3(t)\right], \\
&~\frac{d\beta_4(t)}{dt}=-\frac{1}{\beta}\left[\tfrac{1}{3}\beta^2_0(t)+24\beta_1(t)\beta_4(t)
+12\beta_1(t)\beta_6(t)+40\beta_2(t)\beta_3(t)+10\beta_2(t)\beta_7(t)\right. \notag \\
&~~~~~~~~~~~~~~~~~~~+96\beta_3(t)\beta_4(t)+16\beta_3(t)\beta_5(t)
+48\beta_3(t)\beta_6(t)+24\beta_4(t)\beta_7(t) \notag \\[1mm]
\label{eq:DiffbetaGF4}
&~~~~~~~~~~~~~~~~~~~\left.-~6\beta_2(t)-62\beta_4(t)\right], \\
\label{eq:DiffbetaGF5}
&~\frac{d\beta_5(t)}{dt}=-\frac{1}{\beta}\left[-2\beta^2_0(t)+60\beta_1(t)\beta_2(t)
-56\beta_5(t)-24\beta_6(t)\right], \\
&~\frac{d\beta_6(t)}{dt}=-\frac{1}{\beta}\left[\tfrac{2}{3}\beta^2_0(t)+60\beta_1(t)\beta_4(t)
+60\beta_1(t)\beta_6(t)+100\beta_2(t)\beta_3(t)+240\beta_3(t)\beta_4(t)\right. \notag \\
\label{eq:DiffbetaGF6}
&~~~~~~~~~~~~~~~~~~~\left.+~8\beta_5(t)\beta_7(t)+24\beta_6(t)\beta_7(t)
-24\beta_5(t)-56\beta_6(t)\right], \\
&~\frac{d\beta_7(t)}{dt}=-\frac{1}{\beta}\left[-\tfrac{4}{3}\beta^2_0(t)+60\beta_1(t)\beta_3(t)
+100\beta_2(t)\beta_4(t)+240\beta^2_3(t)+240\beta^2_4(t) \right. \notag \\
\label{eq:DiffbetaGF7}
&~~~~~~~~~~~~~~~~~~~\left.+~32\beta_5(t)\beta_6(t)+48\beta^2_6(t)+12\beta^2_7(t)-72\beta_7(t)\right], 
\end{align}
within our truncation scheme.
These are no longer solved analytically since the equations are quadratic in $\beta_i(t)$,
which is the consequence of the fact that Eq.~\eqref{eq:MasterEqGFEx} is quadratic in $S_t$.

\begin{figure}[tbhp]
\centering
\hspace{2cm}\includegraphics[width=12cm]{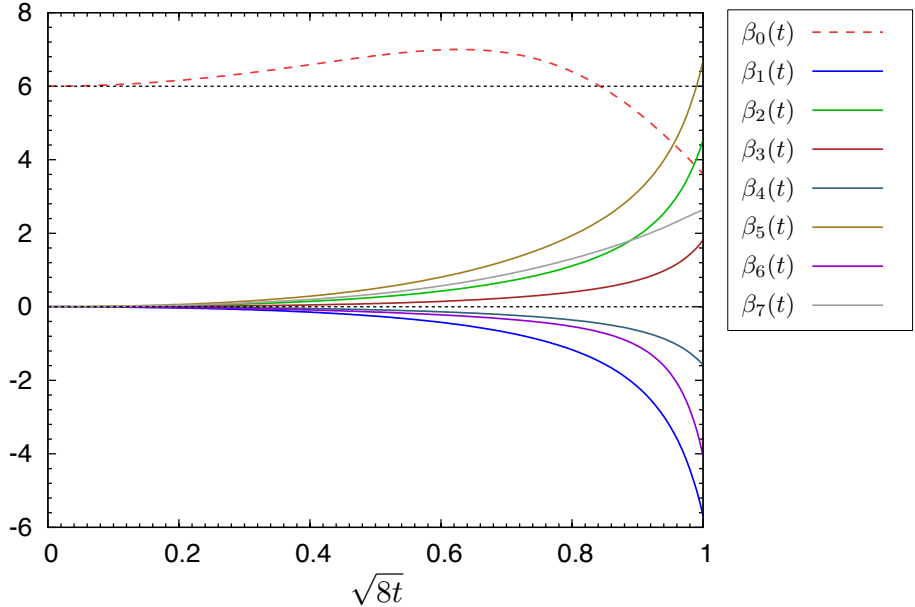}
\caption{Numerical solutions of $\beta_i(t)$ as functions of $\sqrt{8t}$ 
with $\beta=6$ as same as Fig.~\ref{fig:SolWF}.
The dashed line represents $\beta_0(t)$ and the solid lines represent $\beta_{i\geq1}(t)$.}
\label{fig:SolGF}
\end{figure}

As in Sect.~\ref{subsec:SolutionWF},
the numerical solutions of Eqs.~\eqref{eq:DiffbetaGF0}-\eqref{eq:DiffbetaGF7} with $\beta=6$
are plotted in Fig.~\ref{fig:SolGF} as the functions of $\sqrt{8t}$.
The dashed line represents $\beta_0(t)$ and the solid lines represent $\beta_{i\geq1}(t)$.
Fig.~\ref{fig:SolGF} shows that the behavior of $\beta_0(t)$ is similar to 
that of in Sect.~\ref{subsec:SolutionWF}, while $|\beta_{i\geq1}(t)|$ rapidly grow with increasing $t$.
Comparing Eqs.~\eqref{eq:DiffbetaGF0}-\eqref{eq:DiffbetaGF7} 
with Eq.~\eqref{eq:DiffbetaWF}, 
we have linear and quadratic terms in $\beta_{i\geq1}(t)$ 
on the right hand side of Eqs.~\eqref{eq:DiffbetaGF1}-\eqref{eq:DiffbetaGF7}, 
which amplifies the value of $|\beta_{i\geq1}(t)|$. 
Then $\beta_0(t)$ receives the effect of this growth in $|\beta_{i\geq1}(t)|$ and decrease less than its initial value.
These explain such an interesting time evolution of couplings as seen in Fig.~\ref{fig:SolGF}.



\subsection{Similarities and differences compared to the Wilsonian RG flow}
\label{subsec:RelationRG}
As noted in Sect.~\ref{sec:Intro},
our final goal is to construct a scheme of gauge-invariant renormalization group.
Here we discuss on this subject based on the result in Sect.~\ref{subsec:Example}.

Let us assume that the flow time dependence of $\beta_0(t)$ described above is common to any $\beta_i(t)$:
we assume that a coupling $\beta_i(t)$ associated with a class of Wilson loops or their products
whose ``extent'' is roughly around $\sqrt{8t}\simeq n$ becomes relevant at $(n-1)\lesssim\sqrt{8t}\lesssim n$ and 
turns to be irrelevant at $\sqrt{8t}\gtrsim n$.
This implies that Eq.~\eqref{eq:MasterEqGFEx} contains a picture of the renormalization group and 
$(1+\sqrt{8t})a$ could be regarded as a effective lattice spacing.
To confirm this assumption, a truncated theory space should be enlarged
while the number of conceivable classes of Wilson loops increases factorially.

However,
since the gradient flow is just a smearing of the link variables, 
the time evolution of $S_t$ does not have a coarse-graining step.
$S_t$ is defined on the fine lattice even at a large $t$ regime.
Therefore, we need a coarse-graining step to 
eventually construct a RG scheme in this approach,
for example to obtain the beta function of the gauge coupling.
This could be one of future perspectives of this work.
Note that a coarse-graining step on the lattice could define a discretized 
RG equation rather than continuous (differential) one like Eq.~\eqref{eq:MasterEqGF}.

%% file: Summary.tex
In this work, we have first proposed the exact differential equation for the effective action $S_t$
defined by Eq.~\eqref{eq:defEA}, 
which enables us to determine the flow time dependence of the action without lattice simulations.
It is investigated in truncated 8-coupling theory space (Eq.~\eqref{eq:TrunAction} and Fig.~\ref{fig:7WL}) and
the analytical solutions have been obtained (Eq.~\eqref{eq:Sols}).  
Then the equation is extended to more generic cases,
where the flow equation of link variables is not 
limited to the Wilson flow but can be a gradient flow with an arbitrary $t$-dependent action $\mathcal{S}_t$. 
We also have analyzed its concrete example: $\mathcal{S}_t=g^2_0S_t$ in Sect.~\ref{subsec:Example}
and given the solution in the same truncated theory space as the case of the Wilson flow.

For the case of the Wilson flow,
the differential equation of couplings becomes a linear inhomogeneous first-order differential equation
with any truncation, therefore it can be solved analytically.
The solutions are shown in Eq.~\eqref{eq:Sols} and plotted in Fig.~\ref{fig:SolWF}
as the functions of $\sqrt{8t}$.
We have found that the coefficient of the rectangular loop ($\beta_1$)
tends to negative at $t>0$, which agrees with the 
result of Ref.~\cite{Kagimura:2015via} and known improved actions~
\cite{Weisz:1982zw, Weisz:1983bn, Luscher:1984xn, Iwasaki:2011np, Iwasaki:1985we, deForcrand:1999bi}.
However the flow time dependence of couplings is different from that of Ref.~\cite{Kagimura:2015via}.
For our solution, the coefficient of the Wilson plaquette ($\beta_0$) increases at the beginning,
then turns to decrease at around $\sqrt{8t}\simeq1$.

For the case of $\mathcal{S}_t=g^2_0S_t$,
the differential equation of couplings is no longer integrated analytically, since it is quadratic in couplings.
The numerical solutions are shown in Fig.~\ref{fig:SolGF} as same as the case of the Wilson flow.
Similarly to the case of the Wilson flow, $\beta_0(t)$ increases at the beginning then 
decrease less than its initial value.
On the other hand, $|\beta_{i\geq1}(t)|$ look rapidly increasing with the flow time.
These are caused by linear and quadratic terms in $\beta_{i\geq1}(t)$ 
on the right hand side of  Eqs.~\eqref{eq:DiffbetaGF0}-\eqref{eq:DiffbetaGF7},
which are absent in Eq.~\eqref{eq:DiffbetaWF}.
Then we have discussed a possibility to regard this time evolution 
as the Wilsonian RG flow.
For completing our final goal to construct the gauge-invariant RG from the gradient flow,
we need to make the differential equation 
contain a coarse-graining step.

We finally comment on our truncation of the effective action.
The Wilson plaquette and 7 classes of Wilson loops and their products (Fig.~\ref{fig:7WL}), 
which can be constructed by contracting two Wilson plaquettes with a common link variable
has been taken into account in this work.
This truncation is compatible with the small flow time expansion 
only for the case of the Wilson flow in Sect.~\ref{sec:WF}.
Hence, the solution of Eq.~\eqref{eq:MasterEqWF} is exact up to $O(t^{n})$
when all of the classes with contracting up to $n$ Wilson plaquettes are included in $S_t$. 
On the other hand for the case of $\mathcal{S}_t=g^2_0S_t$ in Sect.~\ref{subsec:Example},
we should include other loops in $S_t$ if we want an accurate solution 
in the small flow time expansion.

%% file: Appendix.tex
For evaluating the first term on the right hand side of Eq.~\eqref{eq:MasterEqGFEx},
we need to compute the coefficient $T_{ij,k}~(i,j,k=1,\dots,7)$ defined by
\begin{align}
\sum_{x,\mu}\pa^a_{x,\mu}\mathcal{W}_i\pa^a_{x,\mu}\mathcal{W}_j=
\sum_{k=0}^7 T_{ij,k}\mathcal{W}_k+\cdots,
\end{align}
in addition to Eqs.~\eqref{eq:W0}-\eqref{eq:W7}.
By this definition, $T_{ij,k}$ is symmetric in the exchange between $i$ and $j$: $T_{ij,k}=T_{ji,k}$.
We list non-zero elements of $T_{ij,k}$ in the following,
\begin{equation}
\begin{array}{lllll}
T_{11,1}=12,~&T_{22,1}=28,~&T_{25,1}=20,~&T_{12,2}=24, \\
T_{15,2}=12,~&T_{13,3}=24,~&T_{17,3}=6,~&T_{24,3}=40, \\
T_{26,3}=20,~&T_{33,3}=48,~&T_{37,3}=24,~&T_{44,3}=48, \\
T_{45,3}=16,~&T_{46,3}=48,~&T_{14,4}=24,~&T_{16,4}=12, \\
T_{23,4}=40,~&T_{27,4}=10,~&T_{34,4}=96,~&T_{35,4}=16, \\
T_{36,4}=48,~&T_{47,4}=24,~&T_{12,5}=60,~&T_{14,6}=60, \\
T_{16,6}=60,~&T_{23,6}=100,~&T_{34,6}=240,~&T_{57,6}=8, \\
T_{67,6}=24,~&T_{13,7}=60,~&T_{24,7}=100,~&T_{33,7}=240, \\
T_{44,7}=240,~&T_{56,7}=32,~&T_{66,7}=48,~&T_{77,7}=12.
\end{array}
\end{equation}

For the second term,
the following relations are needed in addition to Eq.~\eqref{eq:W0^2},
\begin{align}
\label{eq:W1^2}
&~\sum_{x,\mu}(\pa^a_{x,\mu})^2\mathcal{W}_1=-8\mathcal{W}_1, \\
\label{eq:W2^2}
&~\sum_{x,\mu}(\pa^a_{x,\mu})^2\mathcal{W}_2=-\tfrac{31}{3}\mathcal{W}_2-\mathcal{W}_4, \\
\label{eq:W3^2}
&~\sum_{x,\mu}(\pa^a_{x,\mu})^2\mathcal{W}_3=-11\mathcal{W}_3+\mathcal{W}_1, \\
\label{eq:W4^2}
&~\sum_{x,\mu}(\pa^a_{x,\mu})^2\mathcal{W}_4=-\tfrac{31}{3}\mathcal{W}_4-\mathcal{W}_2, \\
\label{eq:W5^2}
&~\sum_{x,\mu}(\pa^a_{x,\mu})^2\mathcal{W}_5=-\tfrac{28}{3}\mathcal{W}_5-4\mathcal{W}_6, \\
\label{eq:W6^2}
&~\sum_{x,\mu}(\pa^a_{x,\mu})^2\mathcal{W}_6=-\tfrac{28}{3}\mathcal{W}_6-4\mathcal{W}_5, \\
\label{eq:W7^2}
&~\sum_{x,\mu}(\pa^a_{x,\mu})^2\mathcal{W}_7=-12\mathcal{W}_7+({\rm independent~of~}V).
\end{align}